\documentstyle[11pt,epsf,wrapfig]{article}
\makeatletter
\@addtoreset{equation}{section}
\def\theequation{\arabic{section}\raise.5ex\hbox{.}\arabic{equation}}
\makeatother
\begin{document}
\baselineskip 0.7cm

\setcounter{footnote}{1}

\begin{titlepage}

\begin{flushright}
KUNS-1366 \\
HE(TH)~95/16 \\
hep-ph/9510414
\\
October, 1995
\end{flushright}

\vskip 0.35cm
\begin{center}
{\large \bf 
Vector-like Strong Coupling Theory with Small S and T Parameters
\footnote{
to appear in the Proc. of YKIS 95.
}
}
\vskip 1.2cm
Nobuhiro Maekawa
\footnote
{e-mail address : maekawa@gauge.scphys.kyoto-u.ac.jp
}

\vskip 0.4cm

{\it Department of Physics, Kyoto University,\\
      Kyoto 606-01, Japan}

\vskip 1.5cm

\abstract{
We discuss a mechanism which can reduce the S and T parameters in
dynamical electroweak symmetry breaking scenario. It is
interesting that not only T but also S parameters can be made small
even if large isospin violation exists, which can realize the heavy
top quark.  The point is that if massive
vector-like fermions, which have gauge invariant masses, can condense
and break the electroweak symmetry, the S and T parameters can be
small because of the decoupling theorem.  Moreover, the model predicts a 
heavier Higgs boson, which may be found in LHC. The
presentation basically follows the original papers
\cite{mike}.
}

\vskip 1.5cm

%(to be published in {\sl Phys. Rev.} {\bf D52} (1995) 1684)
\end{center}
\end{titlepage}

%\begin{document}
%\maketitle
\section{Introduction}
Dynamical electroweak symmetry breaking is one of the most attractive
solutions for naturalness problem in Higgs sector
\cite{TC}. 
Unfortunately,
there is no realistic model because of some difficulties. It is hard
to get realistic fermion masses (including heavy top quark ) without
large flavor changing neutral current(FCNC). It is also difficult to
realize the smallness of S and T parameters. These parameters are
defined as
\begin{eqnarray}
{\rm T}&\equiv & {4\pi\over {\rm  sin}^2\theta_W m_W^2}
      \left[\Pi_{11}^{\rm new}(0)-\Pi_{33}^{\rm new}(0)\right] 
      \sim {1\over \alpha}(\rho-\rho_{\rm SM}),\\
{\rm S}&\equiv & 16\pi {d^2\over dq^2}
      \left.\left[\Pi_{33}^{\rm new}-\Pi_{3Q}^{\rm new}\right]
      \right|_{q^2=0},
\end{eqnarray}
where we have adopted the notation of Peskin and Takeuchi
\cite{STU},
and $m_W$ and $\theta_W$ are the mass of the $W$ boson and the
Weinberg angle, respectively. The T parameter is defined as charged 
current contributions
minus neutral current contributions. Therefore this parameter is 
essentially same as $\Delta 
\rho$ parameter except factor $\alpha$, which is sensitive to isospin
violation. The experimental value
\cite{Matsumoto}
 is almost 0, at least less than
1. It is hard to realize large isospin violation in nature, for
example, heavy top quark, keeping the T parameter small in dynamical
electroweak symmetry breaking scenario
\cite{chiv}. The smallness of the S
parameter
severely constrains the dynamical
electroweak symmetry breaking models, for example, the number of new
$SU(2)_L$ doublets
\cite{STU}. 
Since most of dynamical electroweak symmetry
breaking scenario needs some new $SU(2)_L$ doublet fermions, it is not 
so easy to understand the smallness of the S parameter. 

Here
we would like to discuss a mechanism which can reduce the S and T
parameters
\cite{mike}. 
It is interesting that not only S but also T parameter can 
be made small even if large isospin violation exists.

The point is simple and the followings. Peskin-Takeuchi's S and T
parameters are so called ``non-decoupling'' parameters. Actually
particles with large $SU(2)_L\times U(1)_Y$ breaking masses like 4th
generation fermions are not decoupled in a sense. But of course,
particles with large  $SU(2)_L\times U(1)_Y$ invariant masses like
SUSY particles or heavy particles in grand unified theories must be
decoupled because of decoupling theorem
\cite{AC}. 
From this fact, we can expect 
that the S and T parameters can be small if massive vector-like
fermions which have large  $SU(2)_L\times U(1)_Y$ invariant masses, can 
condense and break  $SU(2)_L\times U(1)_Y$ symmetry
\cite{mike,DP,GK,Banks}
. Typical behavior
of their contribution to the S and T parameters
\cite{mike,vectorST,LS,zheng}
 is 
\begin{eqnarray}
{\rm T}\propto {m^4\over m_W^2 M^2},\quad {\rm S}\propto {m^2\over
  M^2}.
\end{eqnarray}
As
notations, small $m$ means  $SU(2)_L\times U(1)_Y$ breaking mass and
capital $M$ means  $SU(2)_L\times U(1)_Y$ breaking mass. We got them by
calculating fermion loops. 
%Zheng
%\cite{zheng}
% also got them independently by using
%low energy effective Lagrangian. 
It is interesting that even if large
isospin violation exists, this mechanism works. Namely, even if the up
type fermion mass $m_U$=1TeV and the down type fermion mass $m_D$=0, taking
the gauge invariant masses $M\sim 10$TeV realize the smallness of T
parameter.

Plan of this paper is as follows. After this introduction, we discuss
``non-decoupling'' parameters, which are rather general concept. In
section 3,
we calculate the S and T parameters in a model with massive
vector-like fermions, and we see that indeed the massive heavy
fermions are decoupled in calculating the S and T parameters. But is
it possible that such a heavy particle($\sim 10$TeV) can condense and
break electroweak symmetry? In section 4, we discuss the massive vector-like
fermions condensation mainly induced by 4-fermi interaction. And in
section 5, we
discuss some models.

\section{Non-decoupling parameters}
In this section we study a general concept of ``non-decoupling''
parameters. 

Most of us would like to know what is new physics beyond the
standard model. In order to know it, we need some signals of new
physics at higher energy scale. But there is an obstacle, which is
called ``decoupling theorem''
\cite{AC}. Namely, if the new physics scale
$\Lambda_{NEW}$ is
much larger than the physical scale $\mu$, all effect we can see is
suppressed by $(\mu/\Lambda)^n(n>0)$.
Even if these suppression factor exists, we can see some
new physics effects, for example, proton decay. But it is interesting
that there are ``non-decoupling'' effects in a sense. The plan of this 
session is followings. After brief review of decoupling theorem, we
discuss non-decoupling effects, which exist in theories with a softly
broken symmetry. Here softly means that a symmetry is broken only by
dimensionful parameters. In the SM, $SU(2)_L\times U(1)_Y$ symmetry is 
broken to $U(1)_{Q}$ only by the vacuum expectation value of the Higgs 
field, so non-decoupling parameters exist. One of them are the S, T,
and U parameters.

\subsection{Decoupling theorem}
We would like to review the decoupling theorem
\cite{AC}
 briefly. We only
discuss it in a following model. The Lagrangian is 
\begin{eqnarray}
{\cal L}={1\over 2}(\partial_\mu\phi)^2-{1\over
  2}m^2\phi^2-{\lambda_3\over 3!}\phi^3-{\lambda_4\over 4!}+\bar \psi
(i\partial_\mu \gamma^\mu-M)\psi+g\bar \psi\psi\phi,
\end{eqnarray}
where $\phi$ is a scalar filed and $\psi$
means a Dirac fermion. If the fermion mass $M$ is much larger than the
scalar mass $m$ and the physical scale $\mu$, we can get the effective
Lagrangian by integrating $\psi$. The effect to the
operators with a dimension(dim) 4+$n$ have a suppression
factor $(\mu/M)^n(n>0)$, on 
the other hand, the effect to the operators with dim$\leq 4$
have no suppression factor but we cannot see the
effects by renormalization at the low energy scale.

\subsection{Non-decoupling effect}
But in theories with a softly broken symmetry, non-decoupling effects
exist in dimension 4 operators which break the symmetry. 
The reason is 
as follows. These  operators, which are induced by loop correction,
should be proportional to the dimensionful
\begin{wraptable}{l}{5cm}
\caption{$U(1)_V\times U(1)_A$ charges.}
\label{table:1}
\begin{center}
\begin{tabular}{lcccc} \hline \hline
          & $h_1$ & $h_2$ & $H_1$ & $H_2$ \\ \hline
$U(1)_V$  &   1   &   1   &   2   &   2   \\
$U(1)_A$  &   1   &   -1  &   1   &   -1  \\ \hline
\end{tabular}
\end{center}
%\begin{figure}
%     \epsfxsize= 5 cm   %or \epsfysize= 3 cm
%           \centerline{\epsffile{figure1.eps}}
%        \caption{A Feynman diagram which induce a breaking operator.}
%        \label{figure:1}
%\end{figure}
\end{wraptable}
 parameters $m_B$
%\cite{symanzik}
 which
 break the
symmetry, because the symmetry is unbroken in the limit that
$m_B\rightarrow 0$. By this fact, in the limit that $m_B\rightarrow
\infty$, the couplings of these operators,
which are dimensionless, do not go to zero generally. 
Namely these parameters do 
not have suppression factor like $1/m_B$. 
As an example, we adopt a
little bit complicate model with a Lagrangian
\begin{eqnarray}
{\cal L}=\sum_{i=1}^2(|\partial_\mu h_i|^2+|\partial_\mu
H_i|^2-m_i^2|h_i|^2-M_i^2|H_i|^2)-V(h_i,H_i),
\end{eqnarray}

\noindent
\begin{wrapfigure}{r}{6cm}
%\begin{figure}
     \epsfxsize= 5 cm   %or \epsfysize= 3 cm
           \centerline{\epsffile{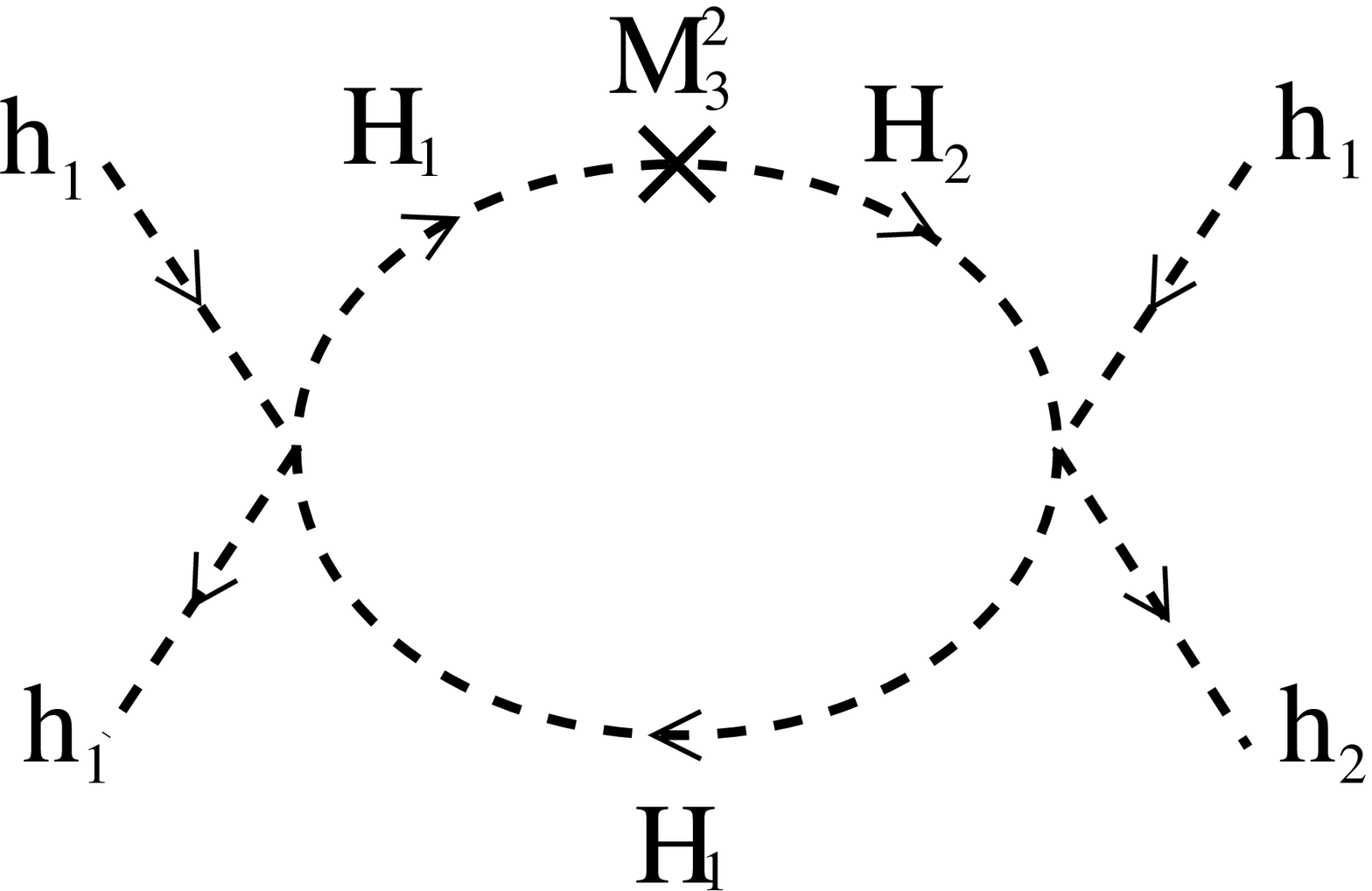}}
        \caption{A Feynman diagram which induce a breaking operator.}
        \label{figure:1}
%\end{figure}
\end{wrapfigure}
where small $h$ mean light scalars fields and
large $H$ mean heavy scalar fields.  
They have quantum numbers under
global symmetries $U(1)_V\times U(1)_A$ as in Table~\ref{table:1}. 
We
 assume that the axial $U(1)_A$ is broken softly by, for example, 
$M_3^2(H_1^\dagger H_2+H_2^\dagger H_1)$ terms. 
Some breaking operators with dim=4
are induced by loop corrections. For example, the operator
$|h_1|^2h_2^\dagger h_1$ is induced 
by a Feynman diagram in Fig.~\ref{figure:1}, which behaves
$M_3^2/f(M_1,M_2,M_3)$.
These effects do not vanish in the limit that $M_1, M_2,M_3\rightarrow 
\infty$. Namely
it seems that heavy fields are not decoupled.

You should notice that this is not a counter example of decoupling
theorem. Integrating the heavy fields $H$, we get the effective
Lagrangian which has no $U(1)_A$ symmetry. In the language of
the effective Lagrangian, the $U(1)_A$ breaking operators like
$|h_1|^2h_2^\dagger h_1$ are
needed in tree level. Once we introduce these operators in tree level,
we cannot see the effect of heavy fields by the renormalization at the low
energy scale. 

We should note that if these
symmetric masses $M_1$ and $M_2$ $\ll$ the breaking masses $M_3$, 
these heavy 
fields are decoupled even in dimension 4 breaking operators.
Namely, in terms of the invariant masses, the heavy fields are
decoupled, but in terms of the breaking masses, the heavy fields are
not decoupled in a sense. Of course, operators with dim>4 
have a suppression factor $1/M^n$ as in the ordinary discussion of
the decoupling theorem. 

As a summary, in theories with a softly broken symmetry like SUSY
models or the standard model(SM), the non-decoupling parameters can be
defined in operators with
dim $\le$ 4, which break the symmetry and
which have no counter terms. 
%As example, in SUSY models, SUSY must be
%broken softly, and in the SM, $SU(2)_L\times U(1)_Y$ symmetry is
%broken only by a dimensionful parameter.

\subsection{Non-decoupling parameters in the standard model}
In the SM, since the symmetry $SU(2)_L\times U(1)_Y$ is broken to $U(1)_Q$
only by a dimensionful parameter(the vacuum expectation value(VEV) of
Higgs field), we can define non-decoupling
parameters following the previous general
discussion. Peskin-Takeuchi's S, T, and U parameters
\cite{STU}
 are one of the 
non-decoupling parameters and they are defined by using the self energy
function of the gauge
bosons $\Pi^{ab}(q^2)(a,b=1,2,3,Q)$. By gauge invariance, 
4 self energy functions $\Pi^{11}=\Pi^{22}$, $\Pi^{33}$, $\Pi^{3Q}$, and
$\Pi^{QQ}$ are
independent. The contribution of the new physics are defined as
$\Pi_{NEW}\equiv \Pi-\Pi_{SM}$, where $\Pi_{SM}$ is from contributions
of the SM.
Since the operators which we are interested in should have
dim $\leq$ 4, we take 1st and 2nd terms in the Tayler
expansion, $\Pi_{NEW}(q^2)=\Pi_{NEW}(0)+q^2\Pi_{NEW}^\prime
(0)+q^2O(q^2/M_{NEW}^2)$.  Since by taking account of unbroken
$U(1)_Q$  symmetry,
$\Pi^{3Q}(0)$ and $\Pi^{QQ}(0)$ are taken to zero, we get the
following 6 parameters
$\Pi_{NEW}^{11}(0)$, $\Pi_{NEW}^{33}(0)$, $\Pi_{NEW}^{11\prime}(0)$,
$\Pi_{NEW}^{33\prime}(0)$, $\Pi_{NEW}^{3Q\prime}(0)$, and 
$\Pi_{NEW}^{QQ\prime}(0)$
 as candidates of non-decoupling parameters.
Moreover since we have counter terms; the VEV of Higgs
field and the two gauge couplings (normalization factors of gauge
fields), we get 3 parameters S, T, and U as
non-decoupling parameters. Indeed, heavy
particles like techni-fermions are not decoupled in a sense. But you
should notice that heavy particles with large $SU(2)_L\times U(1)_Y$
invariant masses must be decoupled.

\section{S and T parameters}
How large $SU(2)_L\times U(1)_Y$ invariant mass $M$ is needed for
small S and T parameters? In this section, we calculate the
contribution to these parameters in a model with vector-like fermions 
which can
have $SU(2)_L\times U(1)_Y$ invariant masses
\cite{vectorST,LS}. 
The Lagrangian of new physics is taken as 
\begin{eqnarray}
{\cal L} &=& \bar Q(i D_\mu \gamma^\mu+M_Q)Q
          +\bar U(i D_\mu \gamma^\mu+M_U)U
          +\bar D(i D_\mu \gamma^\mu+M_D)D   \nonumber \\
         && +(y_U\bar Q_LU_R\phi+y_U^\prime \bar Q_RU_L\phi
          +y_D\bar Q_LD_R\tilde \phi+y_D^\prime \bar Q_RU_L\tilde\phi
          +h.c.).  
\end{eqnarray}
Here $Q=(Q^U,Q^D)$, $U$ and $D$ are Dirac fields which have 
vector-like
couplings to $SU(2)_L\times U(1)_Y$ gauge bosons as
(2,Y), (1,Y+1/2) and (1,Y-1/2) representations, respectively,
and $D_\mu$ are covariant derivatives.
The suffix L and R represent the chirality,
$M_Q$, $M_U$ and $M_D$ are gauge invariant masses,
and $y_U$, $y_U^\prime$, $y_D$ and $y_D^\prime$ are Yukawa
couplings.
%We introduce 3 pairs of
%vector-like fields like this. Namely in addition to ordinary $SU(2)_L$ 
%doublet and singlets, we introduce mirror fermions which have the same 
%quantum charge and opposite chirality. $Q_L$ and $Q_R$ become a Dirac
%field $Q$ and can have gauge invariant mass $M_Q$. And we introduce
%Yukawa couplings like this, here, $\phi$ means Higgs field. 
After
$SU(2)_L\times U(1)_Y$ is broken by the VEV of the Higgs field $\phi$, 
these
fermions have mass matrices,
\begin{eqnarray}
{\cal L}_M&=&
(\bar Q^U_L, \bar U_L) 
\left( \begin{array}{cc} M_Q & m_U \\  m_U^\prime & M_U \end{array} 
\right) 
\left(\begin{array}{c} Q^U_R \\ U_R \end{array} \right)
+(\bar Q^D_L, \bar D_L) 
\left( \begin{array}{cc} M_Q & m_D \\ m_D^\prime & M_D \end{array}
\right)
\left(\begin{array}{c} Q^D_R \\ D_R \end{array} \right)
+h.c. \\
          &=&(\bar U_{L1}, \bar U_{L2}) 
             \left(\begin{array}{cc} m_{U1} & 0 \\ 0 & m_{U2} 
\end{array}\right)
             \left(\begin{array}{c} U_{R1} \\ U_{R2} \end{array} 
\right) 
            +(\bar D_{L1}, \bar D_{L2}) 
             \left(\begin{array}{cc} m_{D1} & 0 \\ 0 & m_{D2} 
\end{array}\right)
             \left(\begin{array}{c} D_{R1} \\ D_{R2} \end{array} 
\right)+h.c. \nonumber
\end{eqnarray}
in which 
\begin{eqnarray}
\left(\begin{array}{c} U_{(L,R)1} \\ U_{(L,R)2} \end{array}\right)=
%\left(\begin{array}{cc} c & -s \\ s & c \end{array}\right)
V_{U(L,R)}\left(\begin{array}{c} Q^U_{(L,R)} \\ U_{(L,R)} \end{array}
 \right),\ 
\left(\begin{array}{c} D_{(L,R)1} \\ D_{(L,R)2} \end{array}\right)=
%\left(\begin{array}{cc} \bar c & -\bar s \\ \bar s & \bar c 
%\end{array}\right)
V_{D(L,R)}\left(\begin{array}{c} Q^D_{(L,R)} \\ D_{(L,R)} \end{array}
 \right), 
\end{eqnarray}     
and $SU(2)_L\times U(1)_Y$ breaking masses $m_U=y_U v$, 
$m_U^\prime=y_U^\prime v$, $m_D=y_D v$ and $m_D^\prime=y_D v$.
Here $V_{(U,D)(L,R)}$ are unitary matrices.
In this paper, as notations,
capital $M$ mean gauge invariant masses and small $m$ mean the
breaking masses. Let us calculate the S and T parameters in the limit 
that the breaking mass of the down type fermion $m_D=0$, which means
that the isospin is maximally violated.

Here we calculate the T parameter by perturbation on the breaking
mass $m=m_U$, which is effective in the limit $M>>m$.
In the 0th order and in the 2nd order, 
the 
\begin{wrapfigure}{r}{6cm}
     \epsfxsize= 5 cm   %or \epsfysize= 3 cm
           \centerline{\epsffile{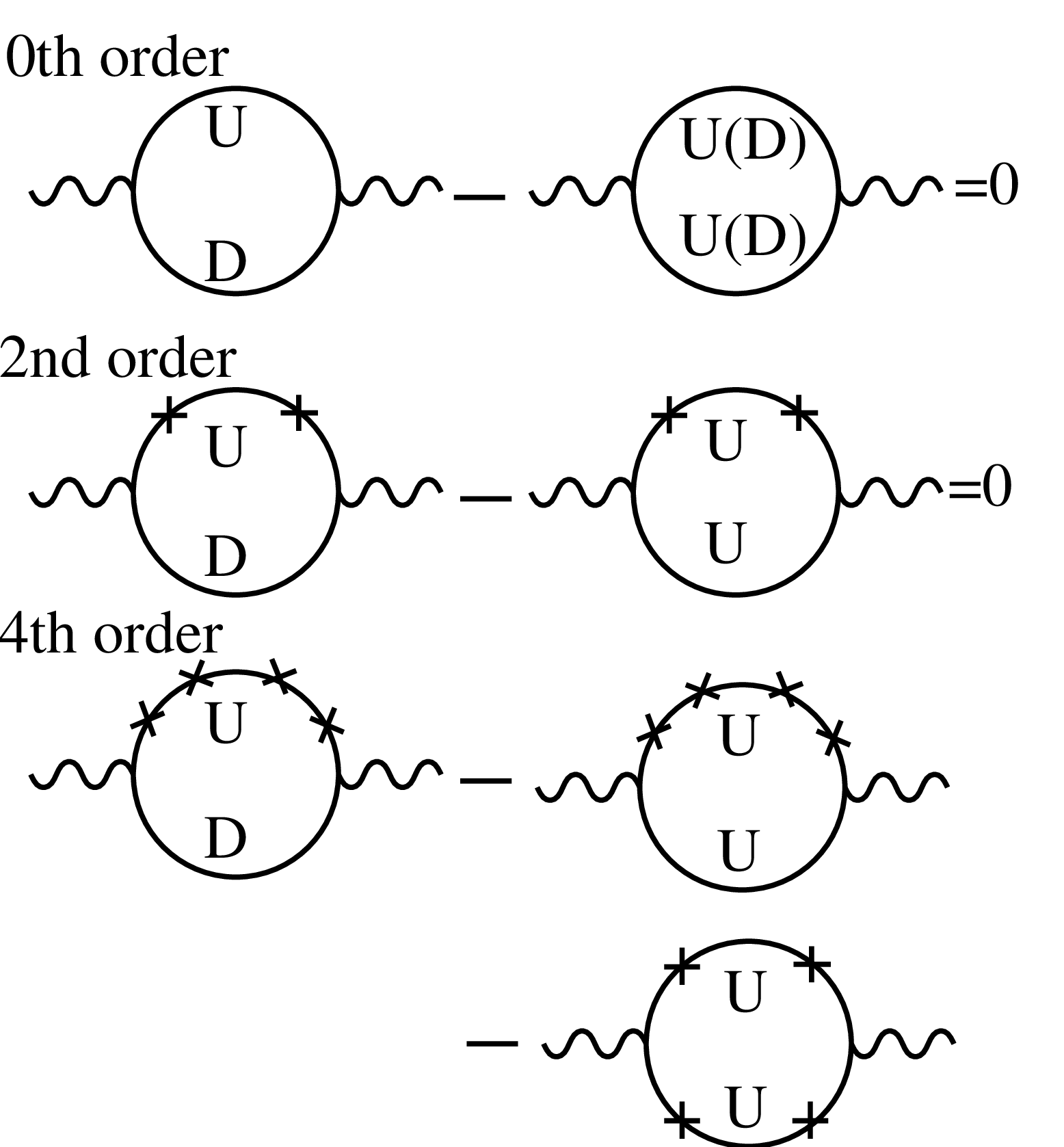}}
        \caption{Feynman diagrams contributing to the T parameter.}
        \label{figure:2}
\end{wrapfigure}
charged current contribution and the neutral
current contribution are cancelled (see
 Fig.~\ref{figure:2}).
And in the 4th order, we can get the
non-vanishing value
\begin{eqnarray*}
%{\rm S}&=&{2N\over 15\pi}\left(m\over M\right)^2
%    +O\left(\left({m\over M}\right)^4\right), \\
{\rm T}&=&{13N m^2\over 480\pi \sin^2\theta_W m_W^2}\left[\left(
m\over M\right)^2
    +O\left(\left({m\over M}\right)^4\right)\right].
\end{eqnarray*}
Even if the down-type fermions have the 
breaking masses, the 2nd order cancellation exists, which guarantees
that these heavy fermions are decoupled in calculating the T
parameter.

Such a decoupling factor can be seen also in calculating the S
parameter,
\begin{eqnarray}
{\rm S}&=&{2N\over 15\pi}\left(m\over M\right)^2
    +O\left(\left({m\over M}\right)^4\right).
%{\rm T}&=&{13N m^2\over 480\pi \sin^2\theta_W m_W^2}\left[\left(
%m\over M\right)^2
%    +O\left(\left({m\over M}\right)^4\right)\right].
\end{eqnarray}
 If we take the breaking mass $m$ = 1 TeV and the
invariant mass $M$ = 10 TeV, we get  S $\sim 0.0004N$ and T $\sim
0.06N$.
It is 
interesting that even if large isospin violation ($m_U=1$TeV, $m_D=0$) 
exists, the T
parameter can be small.

\section{Condensation of massive vector-like fermions by 4-fermi
  interaction} 
Can the situation that such  heavy particles (O(10TeV)) condense and
break electroweak symmetry be realized in dynamical models?

The situation is realized in models in which the Georgi-Kaplan mechanism
\cite{GK,Banks}
 works. They discussed the misalignment of vector-like field 
condensation
in order to avoid the flavor changing neutral current (FCNC) problem. 
This possibility is attractive, but
in this section, we would like to discuss another possibility that heavy
particles with $SU(2)_L\times U(1)_Y$ invariant masses condense
and break  electroweak symmetry.
%in this section we would like to discuss the massive vector-like
%fermions' condensation induced mainly by 4-fermi interaction. 
Here we adopt the following Lagrangian with a 4 fermi coupling $G/N$;
\begin{eqnarray}
{\cal L}_4&=&\bar Q (iD^\mu \gamma_\mu -M_Q) Q+\bar U 
(iD^\mu \gamma_\mu-M_U) U
%             \nonumber \\     
    +{G\over N}(\bar Q_L U_R)(\bar U_R Q_L),
%        +h.c.),
\end{eqnarray}
%in which $M_Q$ and $M_U$ are $SU(2)_L\times U(1)_Y$ invariant masses,
%$D^\mu$ is a covariant derivative of the standard gauge groups,
where 
%${\cal L}_{gauge}$ is kinetic parts of gauge fields and 
we neglected the down-type fermion ($D$) and
every 4-fermi interaction except that in Eq. (4.1)
for simplicity.
Here we only assume the chiral structure of the 4-fermi interaction,
which we will discuss later. 
By auxiliary field method, the
Lagrangian can be rewritten as
\begin{eqnarray}
{\cal L}_Y=\bar Q (iD^\mu \gamma_\mu -M_Q) Q+\bar U 
(iD^\mu \gamma_\mu-M_U) U
      -{N\over G} \phi^\dagger \phi
      +(\bar Q_L U_R \phi+h.c.).
\end{eqnarray}
Integrating these fermion
fields, we can get $1/N$ leading potential
\begin{eqnarray}
V&=&{N\over G}v^2-{N\over 8\pi^2} I+const\sim
 \left\{ \begin{array}{l} {N \over G} v^2 \quad
%{1\over 2}\Lambda^4(1- x_Qx_U) 
%\ln v^2\quad 
                                     (v\rightarrow \infty), \\
              {Nv^2\over G}(1-{f(R_Q^2,R_U^2) G\Lambda^2 \over 8\pi^2})
%                  f(x_Q,x_U) \Lambda^2 v^2+const,
\quad (v\rightarrow 0),
                                 \end{array}\right. \\
I&=&{\Lambda^4\over 2}[ \ln (1+2\alpha+R_Q^2R_U^2)+2\alpha \nonumber \\
 && -(\alpha+\beta)^2\ln {1+\alpha+\beta\over \alpha+\beta} 
    -(\alpha-\beta)^2\ln {1+\alpha-\beta\over \alpha-\beta} ], \\
% &\sim& \left\{ \begin{array}{l} {1\over 2}\Lambda^4(1- R_Q^2R_U^2) 
%\ln v^2\quad 
%                                     (v\rightarrow \infty) \\
%                                f(x_Q,x_U) \Lambda^2 v^2+const,
%\quad (v\rightarrow 0)
%                                 \end{array}\right. \\
R_{(Q,U)}&=&{M_{(Q,U)}\over \Lambda},\quad
%\nonumber \\
\alpha={1\over 2}(R_Q^2+R_U^2+v^2/\Lambda^2),\quad
\beta=\sqrt{\alpha^2-R_Q^2R_U^2},\nonumber \\
f(x,y)&=&{1\over 2(1+x)(1+y)}+{1\over 2}
       -\left\{{x^2\over 2(x-y)}\left[2\ln \left({1+x\over x}\right)
-{1\over 1+x}\right]
         +(x\leftrightarrow y)\right\}, \nonumber
\end{eqnarray}
with 
the VEV of Higgs field $\langle \phi 
\rangle=(v,0)$
and 
the cutoff $\Lambda$.
The behaviors of the potential in $v\rightarrow \infty$ and in $v\sim
0$ in Eq.(4.3) show that 
the critical coupling exists.
\begin{wrapfigure}{r}{6cm}
     \epsfxsize= 6 cm   %or \epsfysize= 3 cm
           \centerline{\epsffile{figure3.eps}}
        \caption{The behavior of the critical coupling.}
        \label{figure:3}
\end{wrapfigure}
If the 4 fermi coupling is smaller
than the critical coupling, $SU(2)_L\times U(1)_Y$ remains unbroken,
and if the 4 fermi coupling is larger than the critical coupling,
$SU(2)_L\times U(1)_Y$ is broken. What is important is that the
situation ($v\ll M$) can be realized by tuning the 4 fermi coupling.
The behavior of the critical coupling is seen in Fig.~\ref{figure:3}.
The ratio of the
critical coupling to that of ordinary Nambu-Jona-Lasinio model depends 
only on $R=M/\Lambda$( we take $M\equiv M_Q=M_U=M_D$ for simplicity). 
This graph
shows that the larger the invariant mass is, the larger critical
coupling is needed for dynamical electroweak symmetry breaking.

Moreover what is interesting is that this model predicts a heavy Higgs 
boson. By integrating fermion fields a la Wilson, we get the effective 
Lagrangian
\begin{eqnarray}
{\cal L}_{eff}&=&\bar Q (iD^\mu \gamma_\mu -M_Q) Q+\bar U 
(iD^\mu \gamma_\mu-M_U) U  \nonumber \\
  &&  +Z_\phi |D_\mu \phi|^2-m_\phi^2 \phi^\dagger \phi
     -{\lambda\over 2}(\phi^\dagger \phi)^2
      +(\bar Q_L U_R \phi+h.c.) +{\cal L}_{gauge},
\end{eqnarray}
where
\begin{eqnarray}
Z_\phi&=&{N\over 16\pi^2}\left( \ln{1+x\over x}-{1\over 1+x}
          -{1\over 3(1+x)^2} \right), \\
\lambda&=&{N\over 8 \pi^2}\left( \ln {1+x\over x}
-{1\over 1+x}
           -{1\over 2(1+x)^2}-{1\over 3(1+x)^3}
           \right),
\end{eqnarray}
from which 
we can calculate the Higgs mass $m_H$ and the breaking mass of up-type 
fermion $m_U$
in $1/N$
leading approximation. Since the Yukawa coupling is normalized 1, 
$m_U=v$.
The
weak boson mass $m_W$ and the Higgs mass are also calculated by ordinary
relations $m_W^2={1\over 2}g_2^2 Z_\phi v^2$ and 
$m_H^2={2\over  Z_\phi}\lambda v^2$. 
From these equations, we can calculate

\noindent
\begin{wrapfigure}{r}{7cm}
%\begin{figure}
     \epsfxsize= 6 cm   %or \epsfysize= 3 cm
           \centerline{\epsffile{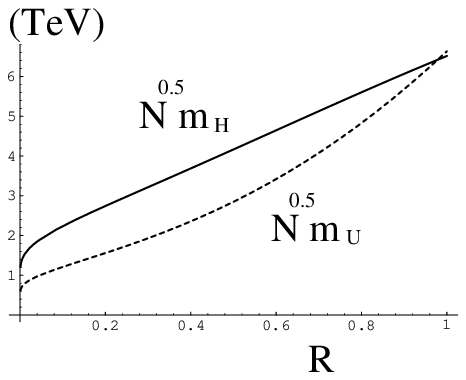}}
        \caption{The Higgs mass $m_H$ and the breaking mass $m_U$ 
          depend only on
          the ratio $R=M/\Lambda$ in $1/N$ leading approximation.}
        \label{figure:4}
%\end{figure}
\end{wrapfigure}
the breaking mass
and the Higgs mass. The behavior of these masses are seen in
Fig.~\ref{figure:4}. The
Higgs mass is O(TeV) in the $1/N$ leading approximation. For example,
if we take $R^2=0.1$, $N=4$, we get $m_U\sim 1$ TeV and $m_H\sim1.5$
TeV. Moreover if we take $M=10$TeV(namely $\Lambda\sim 30$ TeV), the S 
and T parameters become S$\sim 0.0016$ and T$\sim 0.24$.
Note that $\sqrt{N}m_H$ depends only on the ratio $R=M/\Lambda$.
This fact means that the scale of the Higgs mass is decided by the
weak scale, not by the gauge invariant mass scale. Why the heavy
fermions are decoupled in calculating the S and T parameters, and not
in calculating the Higgs mass? This is because the contribution to the 
S and T parameters is finite, on the other hand, the contribution to
the Higgs mass has quadratic divergence.

\section{Some models}
There are some questions and problems. Since we used 4 fermi
interaction, (1) we need some tuning of the coupling in order to realize
the hierarchy $v\ll \Lambda$. (2) What is the origin of the 4 fermi
interaction? We would like to regard the 4 fermi interaction as
effective interaction of some underground physics.
And (3) what is the scale $M$? Since the scale $M$ is
independent of the weak scale and the cut off in this description, we
do not understand why the scale $M$ is near the weak scale or the cut
off scale. (4) How about fermion masses including top quark?

In order to answer these questions, we should build some models.

\subsection{4th family and mirror family}

We can make a dynamical model which is almost same as the SM at the 
low energy scale, if we can use any 4-fermi interactions by hand.
In addition to the strong 4-fermi interaction.
This model is interesting because it does not need so strong a fine 
tuning as top condensation model
\cite{top,BHL}
and because large isospin violation can be realized.
However, we cannot answer the other questions in this model.

\subsection{Techni-fermion(TF) and mirror TF}
Next model can induce the strong 4-fermi interaction by massive gauge
boson exchange, which is similar to the Top color model by Hill
\cite{Hill}. 
We introduce techni-fermions(TFs) $Q_L$, $U_R$, and $D_R$ and
anti-TFs(ATFs) $Q_R$, $U_L$, and $D_L$, which transform as 
$(N,1)$ and $(1,N)$ under gauge groups $SU(N)_G\times SU(N)_{AG}$, 
respectively. Therefore the theory is chiral theory. 
After a scalar field $\Phi_a^b$, which transform as $(N, \bar N)$, 
has VEV $\langle \Phi_a^b \rangle=V\delta_a^b$, the gauge
groups are broken to $SU(N)_V$, and the TFs and the ATFs become
vector-like fermions, which can have gauge invariant masses.

If we take $SU(N)_G\times SU(N)_{AG}$ couplings as $g_G$ 
and 
$g_{AG}$, 
respectively, massless $SU(N)_V$ gauge bosons $V^a_\mu$ with the 
gauge coupling
$g_V$ and massive vector bosons $A^a_\mu$ (`techni-colorons') are 
defined by
\begin{eqnarray}
A^a_\mu=\cos \theta A^a_{G\mu}-\sin \theta A^a_{AG\mu} \quad
V^a_\mu=\sin \theta A^a_{G\mu}+\cos \theta A^a_{AG\mu},
\end{eqnarray}
where $A^a_{G\mu}$ and $A^a_{AG\mu}$ are gauge fields of the groups 
$SU(N)_G$ and $SU(N)_{AG}$, respectively, and
$\tan \theta = g_{AG}/g_G$, $1/g_V^2=1/g_G^2+1/g_{AG}^2$.
The mass of the massive vector fields $A^a_\mu$ is given by
$M_A=\sqrt{g_G^2+g_{AG}^2} V$, which will become the cutoff scale
$\Lambda$ of the low energy vector-like theory.
The currents of $SU(N_{TC})_V$ and $SU(N_{TC})_A$ will be
\begin{eqnarray}
J_{V\mu}^a&=&g_V(\bar Q\gamma_\mu (T^a)Q+\bar U \gamma_\mu (T^a)U
           +\bar D\gamma_\mu(T^a) D) \\
J_{A\mu}^a&=&g_V\cot \theta (\bar Q_L\gamma_\mu (T^a)Q_L
           +\bar U_R\gamma_\mu (T^a)U_R+\bar D_R\gamma_\mu(T^a) D_R)
           \nonumber \\
&&       -g_V\tan \theta (\bar Q_R\gamma_\mu(T^a) Q_R
           +\bar U_L\gamma_\mu(T^a) U_L+\bar D_L\gamma_\mu(T^a) D_L).
\end{eqnarray}
Suppose that $g_G>>g_{AG}$, namely, $\cot \theta>>1$ and
we assume that the gauge coupling $g_G$ at the scale 
$\Lambda$ is
large enough to induce the strong 4-fermi interaction as we discussed
in the previous section
 ( if the phase transition is second order, it is possible to induce
effectively the strong 4-fermi interaction by tuning the parameter
$g_G$\cite{Luty} ). 
Note that $SU(N)_V$ interaction under the scale $\Lambda$ plays little 
role in
breaking the electroweak symmetry ( namely, the composite scale of the
Higgs field is $\Lambda$ ). 
%Since $N_{TC}+3>N_{TC}$, it can be naturally expected
%that $g_G>g_{AG}$, 
Effective 4-fermi interactions $ J_A^{\mu a} J_{A\mu}^a / M_A^2$ are
induced by $SU(N)_A$ gauge boson exchange.
Since $\cot \theta>>1$, the 4-fermi interactions between TFs becomes 
stronger
than between TF and ATF or between ATFs.
Namely, at the scale $\Lambda$, the strong 4 fermi interactions between
an up-type TQ induced
by the $SU(N)_A$ interaction are 
%using Fierz transformation as
\begin{eqnarray}
{g_V^2\cot^2 \theta \over M_A^2}
\sum_a\bar Q_L \gamma_\mu(T^a)Q_L\bar U_R \gamma^\mu(T^a)U_R 
\sim -{N_{TC}g_V^2\cot^2 \theta \over M_A^2}(\bar Q_L U_R 
\bar U_R Q_L),
\end{eqnarray}
where we have used Fierz transformation.  
in which $T^a$ are generators of $SU(N)$. 
Here we take $T^a$ as generators of $SU(N)$ (${\rm
  tr}T^aT^b=\delta_a^b/2$).
The right-hand side in Eq.(5.4) is nothing but the 4 fermi interaction
discussed previously.
Therefore if $g_G(\Lambda)$ is so strong that the induced $SU(N)_A$ 
interaction
can break $SU(2)_L\times U(1)_Y$, the S and T parameters in this model
can be made small when the breaking scale $m_U$ is much smaller than
the gauge invariant mass scale $M$.
In this model, in addition to the above 4-fermi interactions, there 
exist
strong down-type 4-fermi interactions of down-type TQ's.
By these 4-fermi interactions, the down-type TQ (or TL) may 
condense.
For models with fine tuning such as this model, however, 
the small difference between the up and down
sector can induce the large difference between the VEV's of up and 
down type
TF's
\cite{GNJL,critical}, 
which will be understood by the previous discussions about the 
critical coupling.
In this model, $SU(2)_R$ violation is introduced by the difference 
between 
Yukawa couplings $y_U$ and $y_D$ noted below and the difference of the 
hypercharge. By using this criticality, this model can realize the
large isospin violation
in the TFs sector.

If we introduce Yukawa interactions to techni-fermions and 
anti-techni-fermions;
\begin{eqnarray}
y_Q \bar Q_L^a Q_{Rb} \Phi_a^b+y_U \bar U_R^a U_{Lb} \Phi_a^b
+y_D \bar D_R^a D_{Lb} \Phi_a^b+h.c.,
\end{eqnarray}
then techni-fermions have masses of $yV$, which are $SU(2)_L\times 
U(1)_Y$
invariant masses.
Notice that since this theory is chiral theory above the scale
$\Lambda$, it is natural that the gauge invariant masses $M \leq
\Lambda$.

As a summary, this model can explain the origin of the strong 4-fermi
interaction and the gauge invariant masses, but it does not touch the
fermion mass problem. 

\subsection{TFs and ATFs with ordinary fermions}
We would like to induce the ordinary fermion masses.
In addition to ordinary one family extended technicolor (ETC) model
and ETC gauge group $SU(N+3)_{G}$, we introduce ATFs, which have
reverse chirality of TFs, and another gauge group $SU(N)_{AG}$. 
\begin{wrapfigure}{r}{6cm}
     \epsfxsize= 5 cm   %or \epsfysize= 3 cm
           \centerline{\epsffile{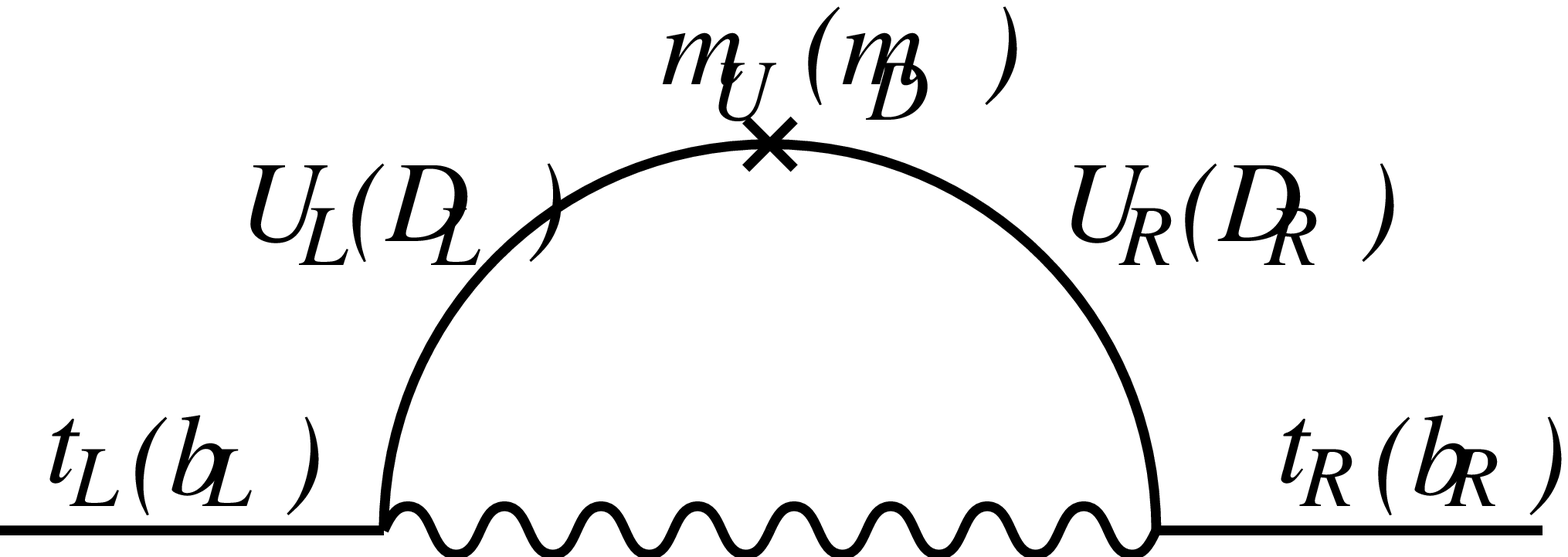}}
        \caption{This Feynman diagram induces the top (bottom) quark
          mass, which are proportional to the breaking mass $m_U$
          ($m_D$).}
        \label{figure:5}
\end{wrapfigure}
As
ordinary one family ETC model, we assume that some other physics ( some 
scalar fields or some strong gauge groups etc.) induce the breaking
pattern $SU(N+3)_G\rightarrow SU(N+2)_G\rightarrow
SU(N+1)_G\rightarrow SU(N)_G$. Moreover, as previously discussed, we 
assume that $SU(N)_G\times SU(N)_{AG}$ is broken to $SU(N)_V$ at scale 
$\Lambda$ and the coupling $g_G$ is enough large to induce the strong
4-fermi interaction. In this model, ordinary matter fields can get
masses from the TFs through the ETC gauge interaction as in
Fig.~\ref{figure:5}
\begin{eqnarray}
%{g_G(\Lambda_3)^2 \over \Lambda_3^2}\bar t t \bar U U
%&\rightarrow& 
m_t\sim 
{g_G(\Lambda_3)^2 \over M_{ETC}^2}\langle\bar U U\rangle
\propto {m_U \Lambda^2\over M_{ETC}^2},\quad
%{g_G(\Lambda_3)^2 \over \Lambda_3^2}\bar b b \bar D D
%&\rightarrow& 
m_b\sim 
{g_G(\Lambda_3)^2 \over \Lambda_3^2}\langle\bar D D\rangle
\propto {m_D \Lambda^2\over M_{ETC}^2}.
\end{eqnarray}
You can easily see that these fermion masses are enhanced because the
dynamics is 

\noindent
\begin{wrapfigure}{r}{6cm}
     \epsfxsize= 5 cm   %or \epsfysize= 3 cm
           \centerline{\epsffile{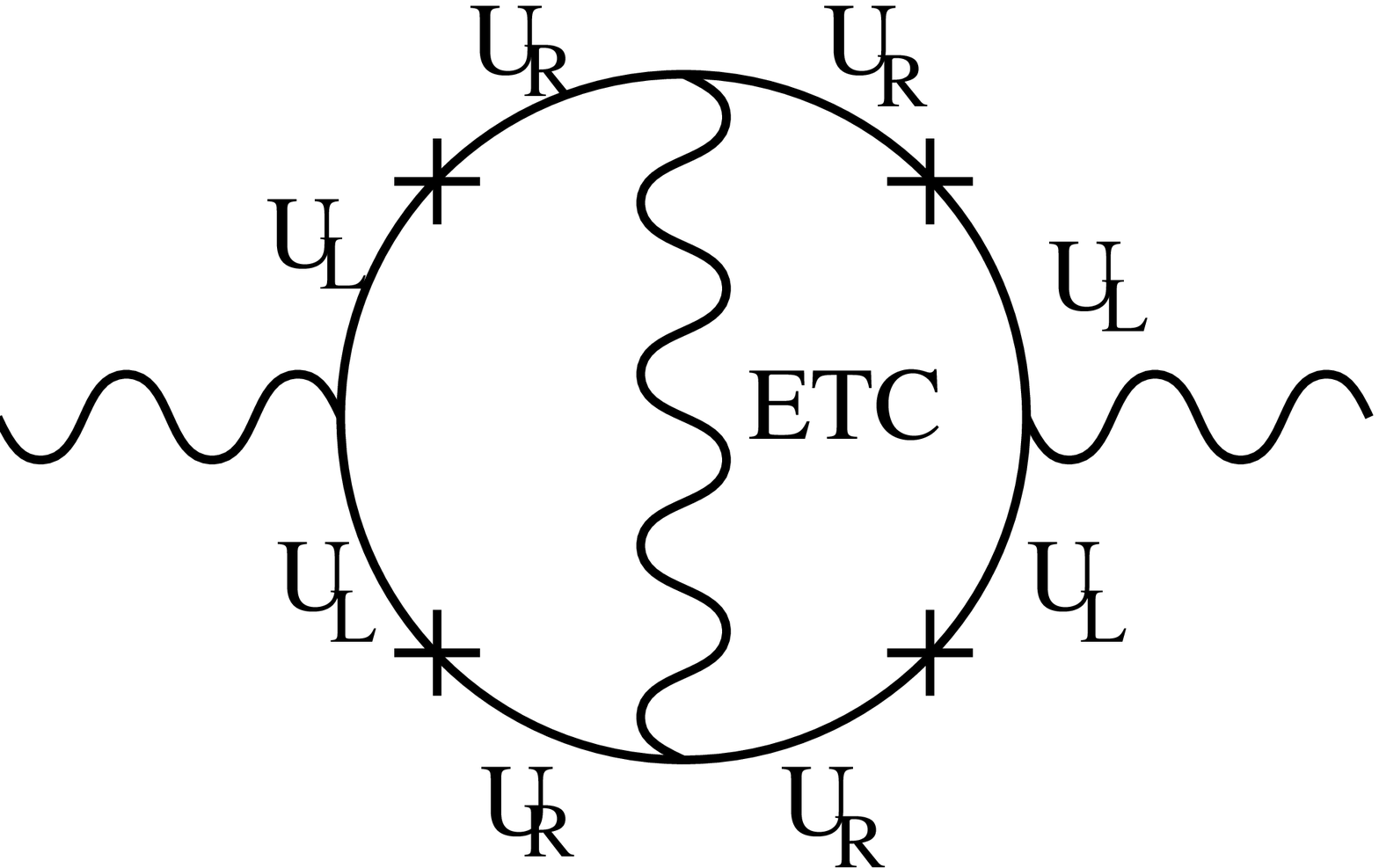}}
        \caption{This contribution to the T parameter will be small.}
        \label{figure:6}
\end{wrapfigure}
almost same as the 4-fermi interaction even if these fermions have the
large gauge invariant masses.
Moreover, you should notice that the large isospin
violation in quark sector ($m_t\gg m_b$) can be induced by large
isospin violation in TF sector ($m_U\gg m_D$).
 
As a comment, though the contribution to the T parameter as in
Fig.~\ref{figure:6} exist in this model, it is small because of the
enhancement of fermion masses.

\section{Discussions and Summary}
Non-decoupling parameters can be defined in theories with a softly
broken symmetry in operators with dim$\leq$ 4 which break the symmetry 
and have no counter term. If fermions with large $SU(2)_L\times
U(1)_Y$ invariant masses condense and break electroweak symmetry, the
S and T parameters can be small. The condensation is possible at least 
by 4-fermi interaction, which can be induced by the massive gauge boson
exchange. A tuning of the coupling is needed, but the tuning enhances
the ordinary fermion masses. The large isospin violation in quark
sector can be induced by large isospin violation in TF sector. A heavy 
Higgs boson ($m_H\sim O({\rm TeV})$) is predicted in $1/N$ leading
approximation. The estimated value seems to be inconsistent with the
triviality bound of the Higgs mass($m_H\leq 530$GeV for $\Lambda\sim$
10 TeV). However, you should remember that $1/N$ subleading effects play 
important roles in calculating the triviality bound. Therefore we can
expect that if we take account of the subleading effects, the
estimated value of the Higgs mass may become lower. LHC may find such
a heavier Higgs boson than that of ordinary SUSY models.

\section{Acknowledgement}
We would like to thank M. Peskin and T. Yanagida for useful
discussions.

\end{document}